\documentclass[11pt]{article}
\usepackage[dvips]{epsfig}
\usepackage[T1]{fontenc}
\usepackage[latin1]{inputenc}
\usepackage{graphicx}
\usepackage[english]{babel}
\usepackage{amsmath}
\usepackage{amssymb}
\usepackage{amsfonts}
\usepackage[T1]{fontenc}
\usepackage{color}
\usepackage{babel}
\usepackage{verbatim}
\usepackage[unicode=true,pdfusetitle,bookmarks=true,bookmarksnumbered=false,bookmarksopen=false,
 breaklinks=false,pdfborder={0 0 1},backref=false,colorlinks=true]{hyperref}
\hypersetup{linkcolor=blue,citecolor=blue}
\makeatletter
\usepackage[dvips]{epsfig}
\usepackage[T1]{fontenc}
\usepackage{color}
\usepackage{pdflscape}
\usepackage{cite}

\begin{document}

\title{\bf Wave Metrics in the Cotton and Conformal Killing Gravity Theories}

\author{Metin G{\" u}rses $^a$\thanks{%
email: gurses@fen.bilkent.edu.tr},~Yaghoub Heydarzade $^a$\thanks{%
email: yheydarzade@bilkent.edu.tr}, and \c{C}etin \c{S}ent{\" u}rk $^b$\thanks{%
email: csenturk@thk.edu.tr}\\\\
\\{\small $^a$Department of Mathematics, Faculty of Sciences, Bilkent University, 06800 Ankara, Turkey}\\
{\small $^b$Department of Aeronautical Engineering,
University of Turkish Aeronautical Association, 06790 Ankara, Turkey}}
\date{\today}

\maketitle
\begin{abstract}

We study wave metrics in the context of Cotton Gravity and Conformal Killing Gravity. First, we consider pp-wave metrics with flat and non-flat wave surfaces and show that they are exact solutions to the field equations of these theories. More explicitly, the field equations reduce to an inhomogeneous Laplace and Helmholtz differential equations, depending on the curvature of the two-dimensional geometry of the wave surfaces. An interesting point here is that the ones with non-flat wave surfaces are not present in classical GR, which manifests a crucial distinction between these theories and GR. Moreover, we investigate Kerr-Schild-Kundt metrics in the context of these theories and show that, from among these metrics, only the AdS wave metrics solve the field equations of these theories. However, AdS spherical and dS hyperbolic wave metrics do not solve the field equations of these theories, which is in contrast to the classical GR. In the case of AdS wave metrics, the field equations of these theories reduce to an inhomogeneous Klein-Gordon equation. We give all the necessary and sufficient conditions for the metric function $V$ to solve these field equations. 

\end{abstract}
\maketitle

\newpage
\tableofcontents
\newpage
\section{Introduction}

Although General Relativity (GR) successfully describes gravity on solar system scales,
some challenges arise in galaxy dynamics, galaxy rotation curves, and the Universe's late time accelerating expansion. To address these issues, researchers are exploring alternative gravity theories. In 2021,  Harada introduced a new geometric theory of gravity \cite{haradacotton}, called
"Cotton gravity" (CG),  gaining attention for its ability to offer new solutions to some problems in GR.  In \cite{haradacotton},  an exact Schwarzschild-like
solution for a static and spherically symmetric source was introduced. It was shown in \cite{haradagalaxy} that the observed rotation curves
of 84 galaxies can be described with CG without need for
the presence of dark matter. Soon after, a general class of spherically symmetric solutions of CG was reported
in \cite{gog, lobo}. The original formulation  of CG necessitates 3rd rank tensorial equations  that  involve the third order derivatives, and hence solving them is more
challenging than solving the field equations in GR. Later, an equivalent  Codazzi formulation of the CG was introduced by Mantica and Molinari \cite{mantica1} where the field equations are reduced to 2nd rank tensorial equations. With this formulation, a self-consistent procedure to generate non-trivial exact solutions in CG that generalize well known GR solutions is introduced in \cite{susman}.
The field equations of CG in the Codazzi formulation has the form
\begin{eqnarray}
&&G_{\mu \nu}=\kappa T_{\mu \nu}+H_{\mu \nu}, \label{cot1}\nonumber\\
&&\nabla_{\alpha}\,{\tilde H}_{\mu \nu}=\nabla_{\mu}\,{\tilde H}_{\alpha \nu}, \label{cot2}
\end{eqnarray}
where $\tilde H_{\mu \nu}=H_{\mu \nu}- Hg_{\mu \nu}$ is a Codazzi tensor with $H=g^{\mu \nu}\,H_{\mu \nu}$. The CG theory was criticized in \cite{barg, haradareply} as having nothing
to add to the content of the standard GR. It was shown in \cite{mantica1} that there is a
rich structure in the theory that includes the standard GR. See also \cite{nuc1,
 sus1, nuc2, sus2} for some issues related to the internal consistency of CG.

 After putting forward the CG theory, Harada introduced yet another theory capable of explaining the observed late time accelerating expansion of the Universe without any need for dark energy  \cite{conf1,conf2}.
The new theory is known as the Conformal Killing Gravity (CKG) and its field equations are also of third order in the derivatives
of the metric tensor. Short after, Mantica and Molinary introduced an equivalent
formulation of the theory in \cite{mant}. In this formulation,  the source of the Einstein's field equations is an energy-momentum tensor augmented
by a divergence-free conformal Killing tensor as
\begin{eqnarray}
&&G_{\mu \nu}=\kappa T_{\mu \nu}+H_{\mu \nu}, \label{con1} \nonumber\\
&&\nabla_{\alpha}\,{\tilde{H}}_{\mu \nu}+\nabla_{\mu}\,{\tilde{H}}_{\nu \alpha}+\nabla_{\nu}\,{\tilde{H}}_{\mu \alpha}=0, \label{conf1}
\end{eqnarray}
where $\tilde{H}_{\mu \nu}=H_{\mu \nu}-\frac{1}{6} H g_{\mu \nu}$ and $H=g^{\mu \nu}\,H_{\mu \nu}$.
CKG satisfies the following theoretical criteria: $(i)$ the cosmological constant is obtained as an integration constant; $(ii)$ the
conservation of the energy-momentum tensor is a consequence of the gravitational field equation, rather than an assumption; and $(iii)$ a conformally flat metric is not necessarily a vacuum solution. The CG and CKG differs except for $(i)$.
The spherical symmetric solution to CKG are studied in \cite{alan, mol, clem}.   In \cite{lobo1, lobo2}, regular black hole and black bounce solutions has been
  reported. The Codazzi formulation of the CKG and its cosmological setup was
discussed in \cite{mant, manti, manti2}. Also, the $pp$-waves in CKG were studied  in \cite{barnes}, and it was reported that the solutions only involve an extra non-propagating term.

The organization of this paper is as follows. First we introduce the generic $pp$-wave metrics and their special cases with flat and non-flat waves surfaces in Section 2. Later we study the ones with flat wave surfaces in Section 3 and the ones with non-flat wave surfaces in Section 4 in the context of both CG and CKG theories. We show that both of these theories admit $pp$-wave metrics with flat and non-flat wave surfaces as exact solutions. It is interesting
that the $pp$-wave metrics with non-flat wave surfaces do not solve the Einstein
field equations with a physical source. After that, we consider the more general class of wave metrics--the Kerr-Schild-Kundt (KSK) metrics--in the context of CG and CKG theories and show that these metrics constitute solutions under some restrictions in Section 5. And finally, we conclude in Section 6. At the end of the paper, there also exists an Appendix which contains the field equations of Cotton and Conformal Killing Gravity theories for $pp$-wave metrics with non-flat wave surfaces.

\section{$pp$-Wave Metrics }

$pp$-Waves, or explicitly \textit{plane-fronted waves with parallel rays}, are well known in GR and defined to be the spacetimes admitting a covariantly constant null vector field $\lambda^\mu$, i.e., the spacetimes with
\begin{equation}
\nabla_\mu\lambda_\nu=0,~~~~\lambda_\mu\lambda^\mu=0.
\end{equation}
In four dimensions, with a suitable coordinate system $x^\mu=(u,v,x^i)$, where $i=1,2$, the line element of the $pp$-waves can be written as \cite{skmhh,rac}
\begin{equation}\label{pp-gen}
ds^2=2 du dv+2 V(u,x^i) du^2+2 A_i(u,x^i) dudx^i+g_{ij}(u,x^i)dx^idx^j,
\end{equation}
where $u$ and $v$ are the double null coordinates, $V(u,x^i)$ is a differentiable function defining the profile of the wave, $A_i(u,x^i)$ is a differentiable function which may carry information about the global properties of the spacetimes, $g_{ij}(u,x^i)$ is a Riemannian metric defined on the two-dimensional wave surfaces, and the null vector appropriately is $\lambda_{\mu}=\delta^u_\mu$. The metric functions $V(u,x^i)$, $A_i(u,x^i)$, and $g_{ij}(u,x^i)$ are all constrained by the field equations of the theory under consideration. In what follows, for simplicity, we shall take $A_i(u,x^i)=0$ and consider two special cases of the generic $pp$-wave metrics (\ref{pp-gen}):

\vspace{0.5cm}
\noindent
1. {\bf $pp$-wave metrics with flat wave surfaces:} In this case, the metric function $g_{ij}(u,x^i)$ is taken to be the flat Euclidean metric $\delta_{ij}$, and the $pp$-wave metric (\ref{pp-gen}) becomes
\begin{equation}\label{pp-flat}
ds^2=2 du dv+2 V(u,x^i) du^2+\delta_{ij}dx^idx^j.
\end{equation}
The Ricci tensor of such metrics is given by \cite{gur1}
\begin{equation}
R_{\mu \nu}=-\rho\, \lambda_{\mu}\, \lambda_{\nu},
\end{equation}
where
\begin{equation}\label{rho}
\rho= \nabla_\perp^2 V ,
\end{equation}
with $\nabla_\perp^2\equiv\partial_i\partial^i$ being the Laplacian defined on the two-dimensional transverse space having the coordinates $x^i$.

\vspace{0.5cm}
\noindent
2. {\bf $pp$-wave metrics with non-flat wave surfaces:} More generally, the metric function $g_{ij}(u,x^i)$ can be assumed to be the metric of a space of constant curvature which we shall denote by $S$ here. This two-dimensional space may be either spherical or hyperbolic depending on whether its curvature constant is positive or negative, respectively. Therefore, the $pp$-wave metric (\ref{pp-gen}), with non-flat wave surfaces, may be written as \cite{heefer}
\begin{equation}\label{pp-nonflat}
ds^2=2 du dv+2 V(u, x^{i})\, du^2+h_{\mu \nu}\, dx^{\mu}\, dx^{\nu}.
\end{equation}
Here $h_{\mu \nu}$ is the four-dimensional metric of a two-dimensional space $S$ with the constant scalar curvature $R$. The Ricci tensor of the spacetime (\ref{pp-nonflat}) is calculated as
\begin{equation}\label{ricci}
R_{\mu \nu}= \frac{R}{2}\, h_{\mu \nu}-\rho\, \lambda_{\mu}\, \lambda_{\nu},
\end{equation}
where
\begin{equation}\label{rho1}
\rho= \nabla_{S}^2 V=h^{\mu \nu}\, \nabla_{\mu} \, \nabla_{\nu}\, V,
\end{equation}
and the covariantly constant null vector $\lambda_{\mu}=\delta_{\mu}^u$ additionally satisfies $\lambda^{\mu}\, h_{\mu \nu}=0$. Here $h^{\mu\alpha}\, h_{\alpha \nu}=\delta^\mu_i \delta^i_\nu $. At this point we should point out that, in obtaining the function $V(u,x^i)$, we need the explicit metric functions of the space $S$ in the case of $pp$-waves with non-flat wave surfaces.

Now let us consider the special $pp$-wave spacetimes (\ref{pp-flat}) and (\ref{pp-nonflat}) in the context of CG and CKG. In what follows, we will assume that there are no external matter fields, i.e., $T_{\mu\nu}=0$.

\section{$pp$-Wave Metrics with Flat Wave Surfaces in Cotton and Conformal Killing Gravity Theories}

$pp$-Wave metrics with flat wave surfaces described in (\ref{pp-flat}) above solve the CG and CKG field equations explicitly. Indeed the solutions can be given respectively as follows.

\vspace{0.2cm}
\noindent
{\bf 1. Cotton Gravity:} From the field equations (\ref{cot1}), we immediately obtain $\tilde{H}_{\mu \nu}=-\rho\, \lambda_{\mu}\, \lambda_{\nu}$ and $\rho=c(u)$, where $c(u)$ is any function of the null coordinate $u$. Hence, thanks to (\ref{rho}), we have
\begin{equation}\label{pp-cot}
\nabla_\perp^2 V=c(u),
\end{equation}
which has the solution
\begin{equation}
V(u,x^i)=V_{0}(u,x^i)+\frac{1}{2}C_{ij}(u)\,x^ix^j,
\end{equation}
where $V_{0}(u,x^i)$ is the solution of the associated homogeneous equation of (\ref{pp-cot}) which corresponds to $pp$-wave solutions of the Einstein equations. Here the trace of the matrix $[C_{ij}(u)]$ is equal to $c(u)$.

\vspace{0.2cm}
\noindent
{\bf 2. Conformal Killing Gravity:} From the field equations (\ref{con1}), we again get $\tilde{H}_{\mu \nu}=-\rho \lambda_{\mu }\, \lambda_{\nu}$, but this time $\rho=c_{0}$ which is a constant. This means that, from (\ref{rho}),
it turns out  \begin{equation}\label{pp-con}
\nabla_\perp^2 V=c_0,
\end{equation}
 and we obtain
\begin{equation}
V(u,x^i)=V_{0}(u,x^i)+\frac{1}{2}C_{ij}(u)\,x^ix^j,
\end{equation}
where $V_{0}(u,x^i)$ is again the solution of the associated homogeneous equation corresponding to $pp$-wave solutions of Einstein's gravity. Here the trace of the matrix $[C_{ij}(u)]$ is equal to $c_{0}$.

In summary, we can say that, for $pp$-wave metrics with flat wave surfaces, both the CG and the CKG field equations reduce to a Laplace equation for $V(u,x^i)$ with source. The homogenous solutions correspond to the solutions of the Einstein equations and the particular solutions are proportional to $x_ix^i$. Thus we have the following theorem.

\vspace{0.3cm}
\noindent
{\bf Theorem 1:} {\it Cotton gravity and Conformal Killing gravity admit $pp$-wave metrics with flat background as solutions with $V=V_{0}+C_{ij}\,x^ix^j$ where $V_{0}$ corresponds to the solutions of the vacuum Einstein field equations and $C_{ij}$ are functions of $u$ in both cases but the trace of the matrix $[C_{ij}(u)]$ is equal to c(u) for the case of Cotton gravity but it is equal to a constant $c_{0}$ for the case of Conformal Killing Gravity Theory.}

\section{$pp$-Wave Metrics with Non-Flat Wave Surfaces in Cotton and Conformal Killing Gravity Theories}

The Ricci tensor and the field equations of the Cotton and Conformal Killing gravity theories are given in more detail in Appendix. We note that $pp$-wave metrics with non-flat wave surfaces do not represent  solutions of the Einstein vacuum field equations unless the scalar curvature is equal to zero in which case these metrics reduce to the standard $pp$-wave metrics with flat wave surfaces. Also, These metrics do not solve the Einstein field equations with a null fluid unless the scalar curvature $R=0$. Hence, when $R \ne 0$, these metrics are not the solutions of Einstein field equations with a reasonable source. Here we shall show that, although $pp$-wave metrics with non-flat wave surfaces do not solve Einstein equations unless $R=0$, they solve both the Cotton and Conformally Killing gravity field equations.

\vspace{0.2cm}
\noindent
{\bf 1. Cotton Gravity:} $pp$-wave metrics with non-flat wave surfaces solve the CG field equations. Indeed, for the metric (\ref{pp-nonflat}), with the help of (\ref{ricci}), we immediately have $\tilde{H}_{\mu \nu}=R_{\mu \nu}+\frac{1}{2} R g_{\mu \nu}=\frac{R}{2} (h_{\mu \nu}+g_{\mu \nu}) -\rho\, \lambda_{\mu}\, \lambda_{\nu}$, where $R$ is the constant scalar curvature. The CG field equations (\ref{cot1}) give that $\rho=-\frac{1}{2}\, R\, V+C(u)$, where $C(u)$ is any function of the null coordinate $u$. Therefore, from (\ref{rho1}), we have
\begin{equation}
\nabla_{S}^2 V+\frac{1}{2} R\, V=C(u).
\end{equation}
This equation is an inhomogeneous Helmholtz equation which can be converted to a homogeneous Helmholtz equation by defining a function $\chi$ such that $\frac{1}{2}\, R\, V=C(u)-\chi$ is satisfied and
\begin{equation}
\nabla_{S}^2 \chi+\frac{1}{2} R\, \chi=0.
\end{equation}
As is obvious, to obtain the profile function $V(u,x^i)$, we need the explicit metric functions of the space $S$.

\vspace{0.2cm}
\noindent
{\bf 2. Conformal Killing Gravity:} For the metric (\ref{pp-nonflat}), using (\ref{ricci}), we get $\tilde{H}_{\mu \nu}=R_{\mu \nu}-\frac{1}{3} R g_{\mu \nu}=\frac{R}{2} (h_{\mu \nu}-\frac{2}{3}g_{\mu \nu}) -\rho\, \lambda_{\mu}\, \lambda_{\nu}$. The field equations (\ref{con1}) give  $\rho= R\, V+ B(u)$, where $B(u)$ is any function of $u$. Then the function $V$ satisfies $ R V=-B(u)+\Psi$, where $\Psi$ satisfies the homogeneous Helmholtz equation
\begin{equation}\label{pp-cot}
\nabla_{S}^2 \Psi- R\, \Psi=0.
\end{equation}
In the limit $R \to 0$, we obtain exactly the same results of the $pp$-waves with flat wave surfaces.

In summary, we can say that, for $pp$-wave metrics with non-flat wave surfaces given in (\ref{pp-nonflat}), both the CG and the CKG field equations reduce to a Helmholtz equation  for $V(u,x^i)$ with source. The homogeneous solutions correspond to the solutions of the Einstein equations. Thus we have the following theorem.

\vspace{0.3cm}
\noindent
{\bf Theorem 2:} {\it Cotton Gravity admits $pp$-wave metrics with non-flat wave surfaces defined in (\ref{pp-nonflat}) as solutions with the profile function $V(u,x^i)$ satisfying the Helmholtz equation
\begin{equation}\label{pp-cot}
\nabla_{S}^2 V+\, (R/2)\, V=C,
\end{equation}
where $C$ is any function of $u$, or letting $\frac{1}{2}\, R\, V=C(u)-\chi$, with a homogeneous Helmholtz equation
\begin{equation}
\nabla_{S}^2 \chi+\frac{1}{2} R\, \chi=0.
\end{equation}
Conformal Killing Gravity field equations reduce also to a similar equation given by
\begin{equation}\label{pp-cot}
\nabla_{S}^2 \Psi-\, R\, \Psi=0,
\end{equation}
where $R V=- B(u)+\Psi(x)$. The homogeneous equation is  the vacuum Einstein field equations.
}

\section{KSK  Metrics in Cotton and Conformal Killing Gravity theories}

Now we shall consider the more general wave metrics, the Kerr-Schild-Kundt (KSK) metrics, which are defined by the line element \cite{gur1,gur2,gur3,gur4,gur5,gur6}
\begin{equation}\label{KSK}
ds^2=ds_{0}^2+2 V (\lambda_{\mu}\,dx^{\mu})^2,
\end{equation}
where $ds_{0}^2$ represents the line element of the maximally symmetric background spacetime in four dimensions. Here $V$ is a differentiable function describing the profile of the wave and $\lambda_{\mu}$ is the null vector having the properties
\begin{eqnarray}
&&\nabla_{\mu}\, \lambda_{\nu}=\frac{1}{2}\, (\lambda_{\mu} \xi_{\nu}+\lambda_{\nu} \xi_{\mu}),\label{xi} \\
&&\lambda^{\mu}\, \lambda_{\mu}=0,\\
&&\lambda^{\mu}\, \partial_{\mu}\,V=0,
\end{eqnarray}
where $\xi^{\mu}$ is a vector satisfying the condition $\lambda^{\mu}\, \xi_{\mu}=0$. The Ricci tensor for the metric (\ref{KSK}) takes the form
\begin{equation}
R_{\mu \nu}= 3K\,g_{\mu \nu}-\rho \lambda_{\mu}\, \lambda_{\nu},
\end{equation}
with $R=12K $, where $K$ is the curvature constant of the background spacetime which may be $K=0$ for the Minkowski, $K>0$ for the de Sitter (dS), or $K<0$ for the anti-de Sitter (AdS) spacetime. Here
\begin{equation}
\rho\equiv\left[\bar{\square}+2 \xi^{\mu}\, \partial_{\mu}+\frac{1}{2} \xi^{\mu}\, \xi_{\mu}+4K \right]\,V
\end{equation}
with $\bar{\square}=g_{0}^{\mu \nu}\, \bar{\nabla}_{\mu}\, \bar{\nabla}_{\nu}$. Then the Einstein tensor takes the form
\begin{equation}
G_{\mu \nu}= -3K\,g_{\mu \nu}-\rho \lambda_{\mu}\, \lambda_{\nu}.
\end{equation}
Therefore, in addition to the pp-wave metrics, depending on type of the background (seed) spacetime, we have the following cases of the KSK spacetimes.

\vspace{0.3cm}
\noindent
{\bf (i) AdS Spherical Wave Metric:} When the background metric is the AdS spacetime, we can take the curvature constant as $K=-\frac{1}{\ell^2}$. Then the metric of the AdS spherical wave spacetime and the corresponding $\xi$ vector defined through (\ref{xi}) are given as follows \cite{gur4}:
\begin{eqnarray}
&&ds^2=\frac{\ell^2}{z^2}\,\left(-dt^2+dx_idx^i +dz^2\right)+2 V(t,x^i,z) (\lambda_{\mu}\, dx^{\mu})^2, \\
&&\lambda_{\mu}= \delta_{\mu}^{t}+\frac{x^i}{r}\delta_{\mu}^{x^i}+\frac{z}{r}\delta_{\mu}^{z},~~\xi_{\mu}=\frac{2}{r}\left( \delta_{\mu}^{t}-\frac{1}{2}\, \lambda_{\mu} \right)+\frac{2}{z}\, \delta_{\mu}^{z}, \label{xi1},
\end{eqnarray}
where $r^2=x_ix^i+z^2$ and $i=1,2$. For the explicit form of the profile function $V(t,x^i,z)$, see \cite{gur4}. Here $z=0$ is a coordinate singularity which is the boundary of the AdS spacetime, hence we assume that $z \ne 0$ in the expressions above and also below. All the curvature invariants are constants.

\vspace{0.3cm}
\noindent
{\bf (ii) AdS Plane Wave Metric:} In the case of AdS plane waves with $K=-\frac{1}{\ell^2}$, the metric and the $\xi$ vector are
\begin{eqnarray}
&&ds^2=\frac{\ell^2}{z^2}\,\left(du^2+2 du dr+dx^2+dz^2 \right)+2 V(u,x,z) du^2, \label{adsmet}\\
&&\lambda_{\mu}=\delta_{\mu}^u,~~\xi_{\mu}=\frac{2}{z}\, \delta_{\mu}^{z}. \label{xi2}
\end{eqnarray}
For details, see \cite{gur1}. In the AdS plane wave spacetime, there exists a null Killing vector filed $\frac{1}{z^2}\, \lambda_{\mu}$.

\vspace{0.3cm}
\noindent
{\bf (ii) dS hyperbolic waves:} If the background metric is the de Sitter spacetime, then we can take $K=\frac{1}{\ell^2}$, and so the dS wave metric and the corresponding $\xi$ vector become \cite{gur5}
\begin{eqnarray}
&&ds^2=\frac{\ell^2}{t^2}\,\left(-dt^2+ dx^2+dy_idy^i\right)+2 V(t,x,y^{i}) (\lambda_{\mu}\, dx^{\mu})^2, \\
&&\lambda_{\mu}=\frac{t}{r}\delta_{\mu}^{t}+\delta_{\mu}^{x}-\frac{y^i}{r}\delta_{\mu}^{y^i},~~~ \xi_{\mu}=\frac{2}{r}\, \left( \delta_{\mu}^{x}-\frac{1}{2} \lambda_{\mu} \right)+\frac{2}{t}\, \delta_{\mu}^{t}, \label{xi3}
\end{eqnarray}
where $r^2=t^2-y_iy^i$ with $i=1,2$. Here $t\neq 0$ and $r\neq 0$.

Now, again assuming $T_{\mu\nu}=0$, we can consider the KSK spacetimes in the context of CG and CKG.

\vspace{0.2cm}
\noindent
{\bf 1. Cotton Gravity: } In the CG case, the field equations (\ref{cot1}) dictate that
\begin{equation}
\tilde{H}_{\mu \nu}=9K\,g_{\mu \nu}-\rho \lambda_{\mu}\, \lambda_{\nu}, \label{heq}
\end{equation}
and
\begin{equation}\label{xiCG}
\xi_{\mu}=\frac{2}{\rho}\left(-\, \partial_{\mu}\, \rho+\psi\, \lambda_{\mu}\right),
\end{equation}
where $\psi$ is any arbitrary function. Comparing (\ref{xiCG}) with the $\xi^{\mu}$ vectors of the AdS spherical wave (\ref{xi1}), AdS plane wave (\ref{xi2}) and the dS hyperbolic wave (\ref{xi3}) metrics, we observe that only the AdS plane wave metric (\ref{adsmet}) is possible. Hence  $\psi=0$  and $\rho =\frac{c_{1}(u)}{z}$, where $c_{1}(u)$ is an arbitrary function of $u$. This implies that the only field equation for Cotton Gravity is given by
\begin{equation}\label{coteqn}
\left[\bar{\square}+ \frac{4z}{\ell^2} \partial_{z}-\frac{2}{\ell^2} \right]\,V=\frac{c_{1}(u)}{z},
\end{equation}
where we have inserted $K=-\frac{1}{\ell^2}$ for the AdS plane wave spacetimes. Explicitly this becomes
\begin{equation}\label{coteqn11}
z^2 (V_{yy}+V_{zz})+2z V_{z}-2 V=\ell^2\,\frac{c_{1}(u)}{z},
\end{equation}
The solution of this equation can be written as $V=V_{h}+V_{p}$ where $V_{h}$ is the solution of the homogeneous equation and $V_{p}$ is the particular solution. It is straightforward to show that $V_{p}=\frac{c_{1}(u) \ell^2}{2 z}$.


\vspace{0.2cm}
\noindent
{\bf 2. Conformal Killing Gravity:} In this case, from the field equations (\ref{con1}), the tensor $\tilde{H}_{\mu \nu}$ takes the form
\begin{equation}
\tilde{H}_{\mu \nu}=-K\,g_{\mu \nu}-\rho \lambda_{\mu}\, \lambda_{\nu}, \label{heq}
\end{equation}
and we obtain
\begin{equation}\label{xiCKG}
\xi_{\mu}=-\frac{1}{2\rho}\, \partial_{\mu}\rho.
\end{equation}
Comparing (\ref{xiCKG}) with (\ref{xi1}), (\ref{xi2}), and (\ref{xi3}), we deduce that the AdS plane wave metric (\ref{adsmet}) with $K=-\frac{1}{\ell^2}$ is the only possible solution and $\rho=\frac{c_{2}}{z^4}$ with $c_{2}$ being an arbitrary constant. Then the only field equation for Conformally Killing Gravity is given by
\begin{equation}\label{coteqn2}
\left[\bar{\square}+ \frac{4z}{\ell^2} \partial_{z}-\frac{2}{\ell^2} \right]\,V=\frac{c_{2}}{z^4},
\end{equation}
which explicitly becomes

\begin{equation}\label{coteqn22}
z^2 (V_{yy}+V_{zz})+2z V_{z}-2 V=\ell^2\,\frac{c_{2}}{z^4}.
\end{equation}
Solution of the above equation can be written as $V=V_{h}+V_{p}$ where $V_{h}$ is the solution of the homogeneous equation and $V_{p}$ is the particular solution. It is straightforward to show that $V_{p}=\frac{c_{2} \ell^2}{10 z^4}$.

\vspace{0.3cm}
\noindent
Now we can state the following theorem.

\vspace{0.3cm}
\noindent
{\bf Theorem 3:} {\it In the KSK metrics, both the Cotton gravity and the Conformal Killing gravity theories admit only the AdS plane wave metrics with $V=V_{h}+V_{p}$. In each case $V_{p}$ corresponds to the particular solutions of the equations (\ref{coteqn11}) and (\ref{coteqn22}) and $V_{h}$ corresponds to the solutions of the vacuum Einstein field equations.}

\section{Conclusion}

We investigated the $pp$-wave metrics in the context of recently introduced Cotton Gravity and Conformal Killing Gravity theories. We showed that both theories admit $pp$-waves with flat and non-flat wave surfaces as solutions. Depending on the geometry of the wave surfaces, the field equations of these theories reduce to an inhomogeneous Laplace equation in the case of flat wave surfaces and Helmholtz equations in the case of non-flat wave surfaces, satisfied by the metric function $V$ defining the profile of the waves. At this point, it should be emphasized that the $pp$-wave metrics with non-flat wave surfaces are not present in classical GR; however, they are solutions here to the CG and CKG due to the higher-derivative nature of the field equations of these theories.

We also considered the more general class of wave metrics so-called Kerr-Schild-Kundt metrics in the context of these theories and we explicitly showed that they admit only the AdS wave metrics as solutions; the others, AdS spherical and dS hyperbolic wave metrics, do not satisfy the field equations of these theories. This is another point that distinguishes these theories from classical GR which admits AdS spherical and dS hyperbolic waves as exact solutions as well. For AdS wave metrics, the field equations of these theories reduce to an inhomogeneous Klein-Gordon equation satisfied by the metric function $V$.

In addition, to give more intuition into internal dynamics of CG and CKG theories, we studied the colliding gravitational plane waves problem in these theories. Since the problem is very complicated due to the third-order derivatives of the field equations of these theories we leave this problem as our future work.

\vspace{2cm}
\section*{Appendix}

In this appendix, we give more information about the $pp$-wave metrics with non-flat wave surfaces in four dimensions. The Ricci tensor of the spacetime (\ref{pp-nonflat}) is computed as
\begin{equation}
R_{\mu \nu}= \frac{R}{2}\, h_{\mu \nu}-\rho\, \lambda_{\mu}\, \lambda_{\nu},
\end{equation}
with
\begin{equation}\label{rho2}
\rho= \nabla_{S}^2 V=h^{\mu \nu}\, \nabla_{\mu} \, \nabla_{\nu}\, V,
\end{equation}
where $h^{\mu\alpha}\, h_{\alpha \nu}=\delta^\mu_i \delta^i_\nu $. The scalar curvature $R$ is assumed to be a constant and the covariantly constant null vector $\lambda_{\mu}=\delta_{\mu}^u$ satisfies $\lambda^{\mu}\, h_{\mu \nu}=0$. In (\ref{rho2}), $\nabla_{S}^2$ is the Laplace-Beltrami operator defined on the non-flat wave surface $S$. The covariant derivative of the Ricci tensor is given as
\begin{equation}
\nabla_{\alpha} R_{\mu \nu}=\frac{R}{2} \left( \delta^{i}_{\nu}\, V_{,i}\, \lambda_{\mu}+\delta^{i}_{\mu}\, V_{,i}\, \lambda_{\nu} \right)\, \lambda_{\alpha}-\rho_{, \alpha}\, \lambda_{\mu}\, \lambda_{\nu},
\end{equation}
where the comma denotes partial derivatives. Since the Ricci scalar $R$ is constant, both the Cotton and Conformal Killing Gravity field equations can be expressed only in terms of the Ricci tensor. Hence we obtain the following equations.

\vspace{0.5cm}
\noindent
{\bf 1. Cotton Gravity Field Equations:}
From (\ref{cot1}), we get the field equations as
\begin{equation}
\nabla_{\alpha}\,R_{\mu \nu}-\nabla_{\mu}\,R_{\alpha              \nu}=\left(\rho_{\mu}+\frac{R}{2}\, \delta^{i}_{\mu}\, V_{,i} \right)\, \lambda_{\nu}\, \lambda_{\alpha}-\left(\rho_{\alpha}+\frac{R}{2}\, \delta^{i}_{\alpha}\, V_{,i} \right)\, \lambda_{\mu}\, \lambda_{\nu}=0.
\end{equation}
These field equations reduce to a more simpler form
\begin{equation}
\rho_{,\alpha}+\frac{R}{2}\, \delta^{i}_{\alpha}\, V_{,i} =\Phi \lambda_{\alpha},
\end{equation}
where $\psi$ is an arbitrary function of the coordinates. Then we obtain
\begin{eqnarray}
&&\rho_{,u}=\psi,\\
&&\rho+\frac{R}{2}\, V=C(u),
\end{eqnarray}
where $C(u)$ is an arbitrary function of $u$ and $\psi$ must satisfy
\begin{equation}
\Phi=C_{,u}-\frac{R}{2}V_{,u}.
\end{equation}

\vspace{0.5cm}
\noindent
{\bf 2. Conformal Killing Gravity Field Equations:}
From (\ref{conf1}), one can write the field equations as
\begin{eqnarray}
&&\nabla_{\alpha}\,R_{\mu \nu}+\nabla_{\mu}\,R_{\nu \alpha}+\nabla_{\nu}\, R_{\mu \alpha} \nonumber \\
&&~~~~~~~~~~=-\left(\rho_{, \alpha}-R\, \delta^{i}_{\alpha}\, V_{,i}\, \right)\, \lambda_{\mu}\, \lambda_{\nu}-\left(\rho_{,\mu}-R\, \delta^{i}_{\mu}\, V_{,i}\, \right)\, \lambda_{\alpha}\, \lambda_{\nu}-\left(\rho_{,\nu}-R\,\delta^{i}_{\nu}\, V_{,i}\, \right)\, \lambda_{\mu}\, \lambda_{\alpha}=0,
\end{eqnarray}
which reduce to
\begin{equation}
\rho_{, \alpha}-R\, \delta^{i}_{\alpha}\, V_{,i}=0.
\end{equation}
This equation can immediately be integrated to obtain
\begin{eqnarray}
&&V(u,x^i)=\frac{-B(u)+\Psi(x^i)}{R},\\
&&\rho=\Psi(x^i),
\end{eqnarray}
where $B$ and $\Psi$ are arbitrary functions and $i=1,2$. Or, equivalently we have
\begin{equation}
\nabla_S^2 \Psi- R \Psi=0.
\end{equation}

\end{document}